# Goldbach Circles and Balloons and Their Cross Correlation

Krishnama Raju Kanchu and Subhash Kak

**Abstract.** Goldbach partitions can be used in creation of ellipses and circles on the number line. We extend this work and determine the count and other properties of concentric Goldbach circles for different values of n. The autocorrelation function of this sequence with respect to even and odd values suggests that it has excellent randomness properties. Cross correlation properties of ellipse and circle sequences are provided that indicate that these sequences have minimal dependencies and, therefore, they can be used in spread spectrum and other cryptographic applications.

## INTRODUCTION

Pseudorandom sequences have applications in cryptography and in spread-spectrum systems [1],[2]. The critical properties governing the cross correlation performance of a good pseudorandom sequence are its RMS and peak values. These sequences have highly peaked autocorrelation function and very small cross correlation. For a good autocorrelation function the ratio of the peak value to the modulus of the highest sidelobe should be as large as possible and cross correlation should be close to zero. Often there is an inverse relationship between the quality of autocorrelation and cross correlation functions for a family of sequences. The optimization of both these functions is, therefore, an important problem in signal theory.

Decimal sequences [3]-[6] provide cross correlation that is zero everywhere for a period that is a product of the periods of the two sequences but because the second half of the maximal-length decimal sequence can be obtained from the first half they cannot be used in all applications. We know other methods to generate sequences with reasonably good cross correlation properties [7],[8] but this is achieved with some loss in the quality of the autocorrelation function. There are also larger issues related to the physics of generation of random sequences that will not be considered here [9].

In recent studies, we have examined several number-theoretic functions to determine new candidates for excellent pseudorandom sequences and these include Pythagorean triples [10] and Goldbach sequences. Specifically, we have considered both Goldbach partitions into prime numbers of even numbers [11],[12] and prime partitions of odd numbers [13]. It is the number of partitions with respect to the integers that can be converted into a binary random sequence. We have shown that these binary sequences have excellent autocorrelation properties. In this paper we first extend this work to the properties of concentric circles and then consider cross correlation properties of various kinds of Goldbach sequences. We show this cross correlation is excellent and so these sequences can be used in applications of cryptography and spread-spectrum systems.

# ELLIPSES AND CIRCLES

In earlier papers [11],[12] we described a method to generate different random sequences that we called circle and ellipse sequences. For the even number *n* the radius *r* is the smallest number so that *n-r* and *n+r* are both primes. Thus for n=14 the radius is 3 since the nearest equidistant primes are 11 and 17 that are equally separated from 14. For the number *n*, the ellipse *(s,t)* exists if *n-s* and *n+t* are primes.

The circle sequence associated with a set of even numbers is the corresponding values of the radii. The ellipse sequence *(1,k)* is the set of random numbers *m* associated with the given natural numbers *n* so that *n-m* and *n+km* are primes. These numbers are reduced to 1 and -1 by computing their mod 4 value.

**Example 1**. Table 1 has the circle sequence for numbers between 30 and 64. The r numbers are the random sequence and they can only be odd numbers. These may be converted to binary sequence by the mapping r mod 4.

**Table 1.** Circle sequence of random numbers, r, for n between 30 and 64.

| *n* | 30 | 32 | 34 | 36 | 38 | 40 | 42 | 44 | 46 | 48 | 50 | 52 | 54 | 56 | 58 | 60 | 62 | 64 |
|---|---|---|---|---|---|---|---|---|---|---|---|---|---|---|---|---|---|---|
| *r* | 1 | 9 | 3 | 5 | 9 | 3 | 1 | 3 | 15 | 5 | 3 | 9 | 7 | 3 | 15 | 1 | 9 | 3 |
| *r* mod *4* | 1 | 1 | -1 | 1 | 1 | -1 | 1 | -1 | -1 | 1 | -1 | 1 | -1 | -1 | -1 | 1 | 1 | -1 |

**Example 2**. The ellipse sequence generated for k = 5, where n ranges from 16 to 56 is shown in Table 2. Note that multiples of k=5 do not figure in the list of n for ellipse sequences for such values will not exist. As explanation, the value for n=16 is 3 since 16-3=13 and 16+3×5=31 are primes.

**Table 2.** Ellipse sequence for *k=5*

| *n* | 16 | 18 | 22 | 24 | 26 | 28 | 32 | 34 | 36 | 38 | 42 | 44 | 46 | 48 | 52 | 54 | 56 |
|---|---|---|---|---|---|---|---|---|---|---|---|---|---|---|---|---|---|
| *m* | 3 | 1 | 3 | 1 | 3 | 5 | 1 | 5 | 5 | 1 | 1 | 3 | 3 | 1 | 5 | 7 | 3 |
| *m* mod *4* | -1 | 1 | -1 | 1 | -1 | 1 | 1 | 1 | 1 | 1 | 1 | -1 | -1 | 1 | 1 | -1 | -1 |

It is clear that by proper choice of the starting point of the circle sequence and by doing this as well as picking values of *k*, any number of Goldbach random sequences can be generated.

# CONCENTRIC CIRCLES

We now describe properties of the concentric circles that are defined for each value of n. In the previous work, the circle chosen for a given value of n was the one with the least radius. Here we wish to list all possible circles associated with a number.

Figure 1 gives a graphical example of concentric circles for small values of n. In particular, one can see the two concentric circles about n=8 and 10, and 12, although the larger ones have been shown somewhat squished.

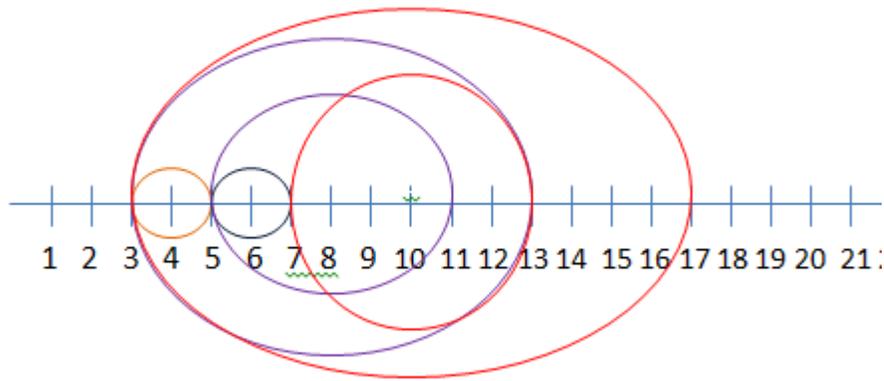

**Figure 1.** Concentric circles around n=4, 6, 8, and 10

Let the count of concentric circles for the number n be given by k(n). Tables 3 and 4 provide some examples of k(n). Note that what we called the circle sequence in the previous section represented merely the first circle information for each of the number points.

**Table 3.** Number of concentric circles k(n), n from 4 to 36

| n | 4 | 6 | 8 | 10 | 12 | 14 | 16 | 18 | 20 | 22 | 24 | 26 | 28 | 30 | 32 | 34 | 36 |
|---|---|---|---|----|----|----|----|----|----|----|----|----|----|----|----|----|----|
| k(n) | 1 | 1 | 2 | 2 | 3 | 2 | 2 | 4 | 3 | 3 | 5 | 3 | 3 | 6 | 5 | 2 | 6 |
| k(n) mod 2 | 1 | 1 | 0 | 0 | 1 | 0 | 0 | 0 | 1 | 1 | 1 | 1 | 1 | 0 | 1 | 0 | 0 |

There are two circles for n=10, namely (7,13) and (3,17), and, likewise, for n=24, the five circles are (5,43), (7,41), (11,37), (17,31), (19,29).

**Table 4.** Number of concentric circles k(n), n from 202 to 226

| n | 202 | 204 | 206 | 208 | 210 | 212 | 214 | 216 | 218 | 220 | 222 | 224 | 226 |
|---|-----|-----|-----|-----|-----|-----|-----|-----|-----|-----|-----|-----|-----|
| k(n) | 11 | 20 | 11 | 10 | 30 | 12 | 9 | 19 | 11 | 14 | 21 | 13 | 12 |
| k(n) mod 2 | 1 | 0 | 1 | 0 | 0 | 0 | 1 | 1 | 1 | 0 | 1 | 1 | 0 |

Figure 2 presents the concentric circle count for small values of n. As in the case of the Goldbach partition count [11], the count k(n) has peaks at multiples of 210. This is to be expected since a number for which there are many partitions will also be expected to have many circles associated with it. As is the case for the Goldbach partition count *g(n)*, the concentric circle count *k(n)* has a local peak when the number of distinct factors of *n* is large. We thus get peaks at multiples of 2×3=6; 2×3×5=30; 2×3×5×7=210; 2×3×5×11=330; 2×3×5×13=390; 2×3×5×7×11=2310; 2×3×5×7×11×13=30030; 2×3×5×7×11×13×17=510510; 2×3×5×7×11×13×17×19=9699690, etc.

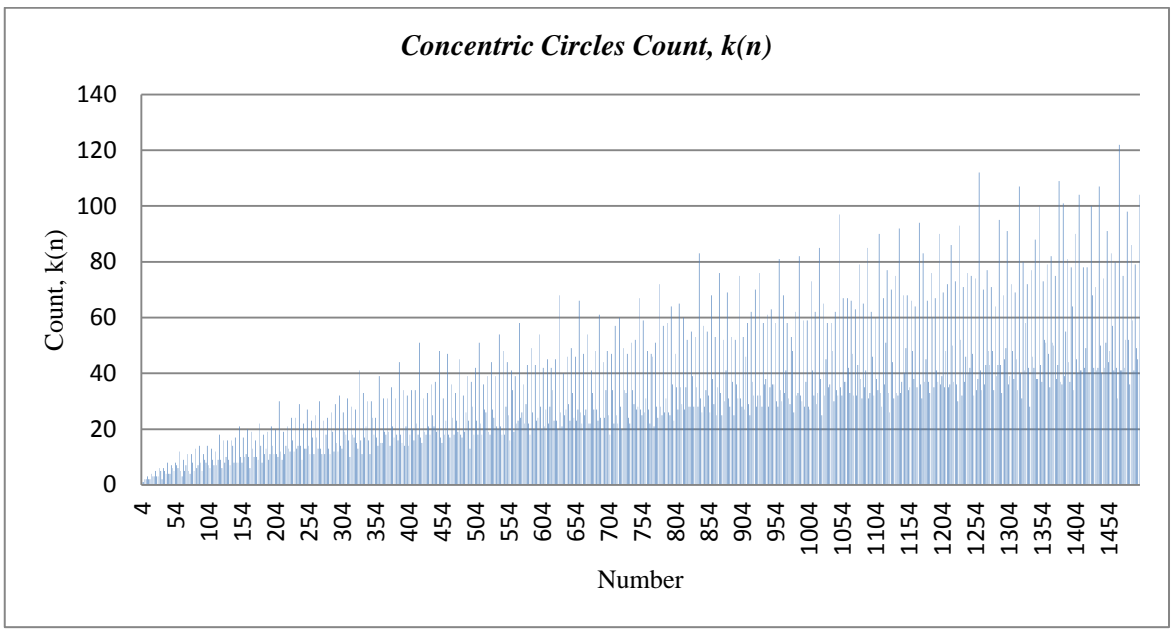

**Figure 2.** Concentric circles count, k(n)

Due to these peaks, we get other conditions such as:

*g(6k) > g(6k+2)*

*g(30k) > g(30k+2)*

*g(210k) > g(210k+2)*

and so on…

To determine the autocorrelation behavior of the concentric circle count sequence k(n), we converted this sequence into a binary sequence based on odd-even parity. The autocorrelation function of this binary sequence is presented in Figure 3. This function satisfies the property of of being approximately two-valued. The maximum value of autocorrelation function for non-zero argument for this example is 0.09, which is quite small.

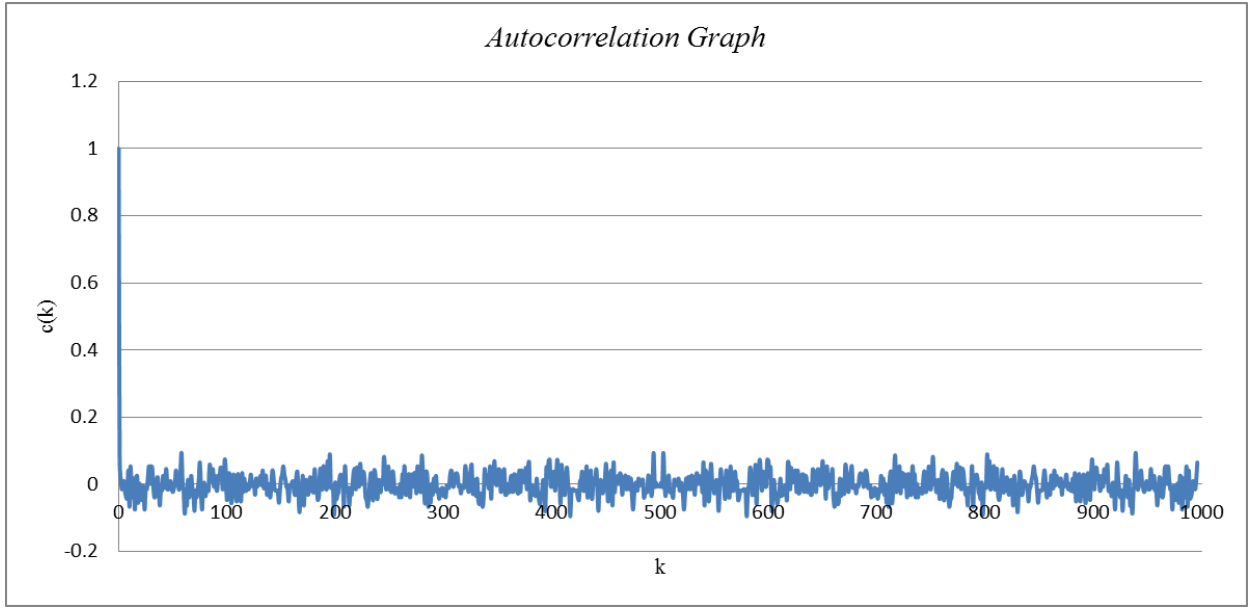

**Figure 3**. Autocorrelation of the binary concentric circle count sequence

## ELLIPSES AND CIRCLES IN TWO DIMENSIONS

Computation of cross correlation requires consideration of two sequences. Figure 4 is a representation of such sequences in two dimensions. We call these circles and ellipses Goldbach balloons.

The cross correlation of two periodic strings, $\{a_i\}$ and $\{b_i\}$, each with period *n*, is given by the following formula:

$$C_{b,a}(k) = 1/n \sum_{k=0}^{n-1} b_i * a_{i+k}$$

The cross correlation between the two binary strings is computed by considering them in the *(1,-1)* form, which is how our sequences are described. The cross correlation function depends on the value of n and $C_{b,a} \neq C_{a,b}$.

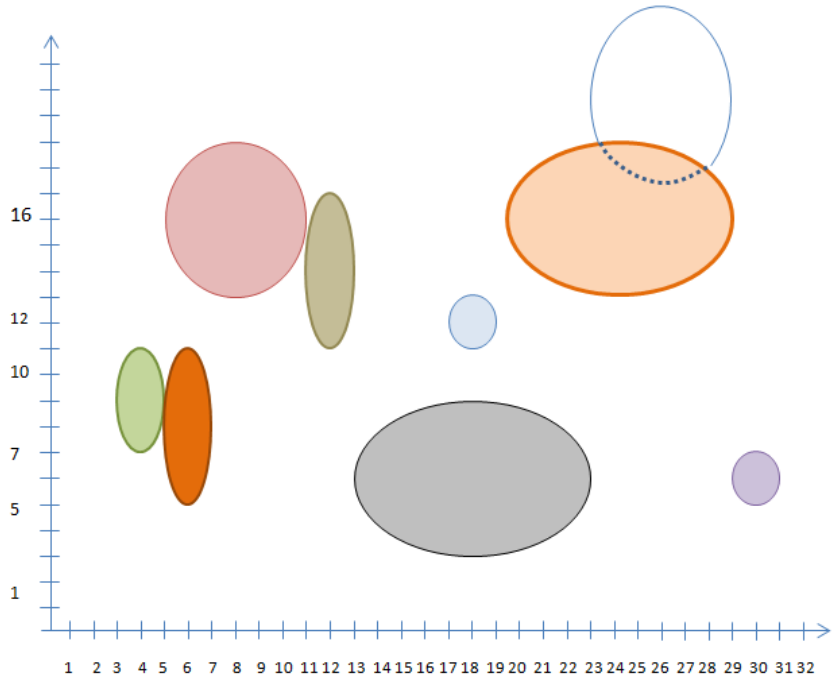

**Figure 4.** Goldbach balloons: circles and ellipses in two dimensions

The cross correlation between the two strings is as shown in Figure 5. We find that in a sequence of length nearly 1,700, the cross correlation value is about 7 percent.

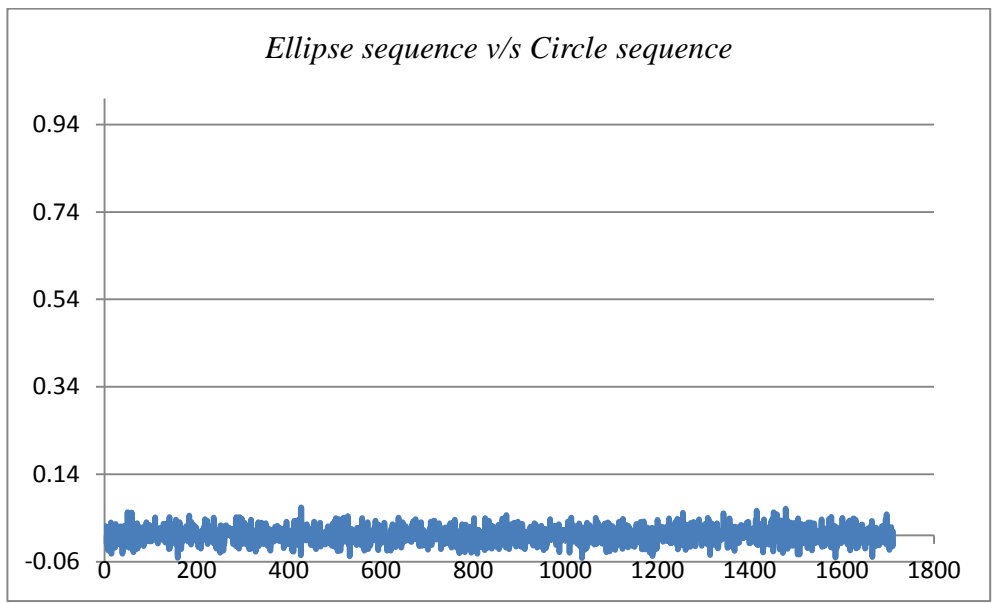

**Figure 5**. Graph showing the cross correlation between ellipse and circle

**Table 5**. Variation in maximum of |C$_{a,b}$(k)| with respect to n

| n | Range |
|---|---|
| 10 | 0.4 |
| 50 | 0.411 |
| 100 | 0.3267 |
| 250 | 0.1784 |
| 500 | 0.1497 |
| 750 | 0.1318 |
| 1000 | 0.1248 |
| 1250 | 0.0983 |
| 1500 | 0.1045 |
| 1750 | 0.0725 |
| 2000 | 0.0764 |
| 2500 | 0.0715 |
| 3000 | 0.0583 |
| 3500 | 0.0591 |
| 4000 | 0.0605 |
| 4500 | 0.0567 |
| 5000 | 0.0486 |

Table 5 shows how the data dependencies of the two strings decrease as the length increases.

In Figure 6 we consider the cross correlation between two ellipse sequences for k=5 and k=7 of length 5000. As the number of points considered in the sequence goes from 10 to 5,000, the value of the peak cross correlation point goes down from a bit over 0.4 to 0.0486.

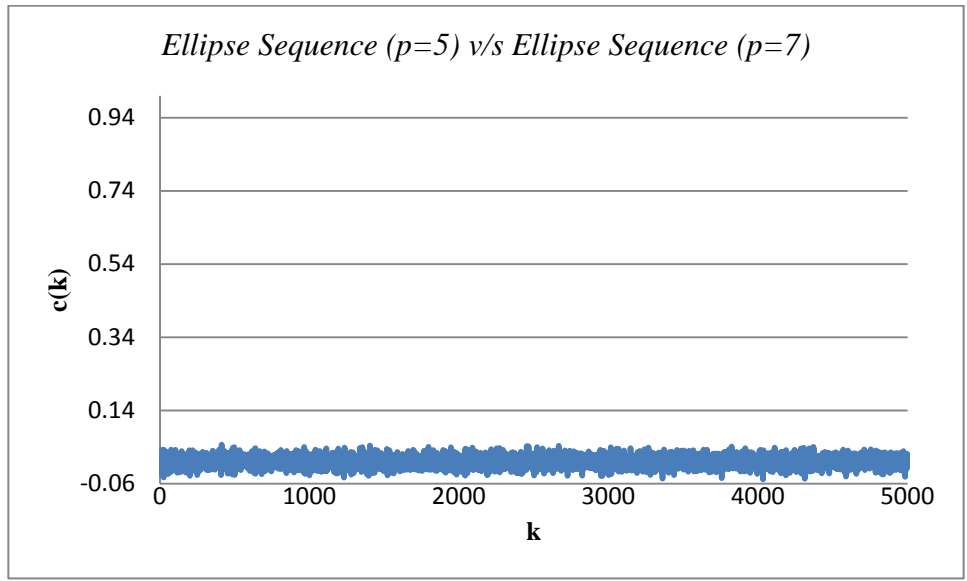

**Figure 6**. Cross correlation between ellipse, p=5 on ellipse, p=7

Figure 7 presents the cross correlation between concentric and Goldbach circle sequences.

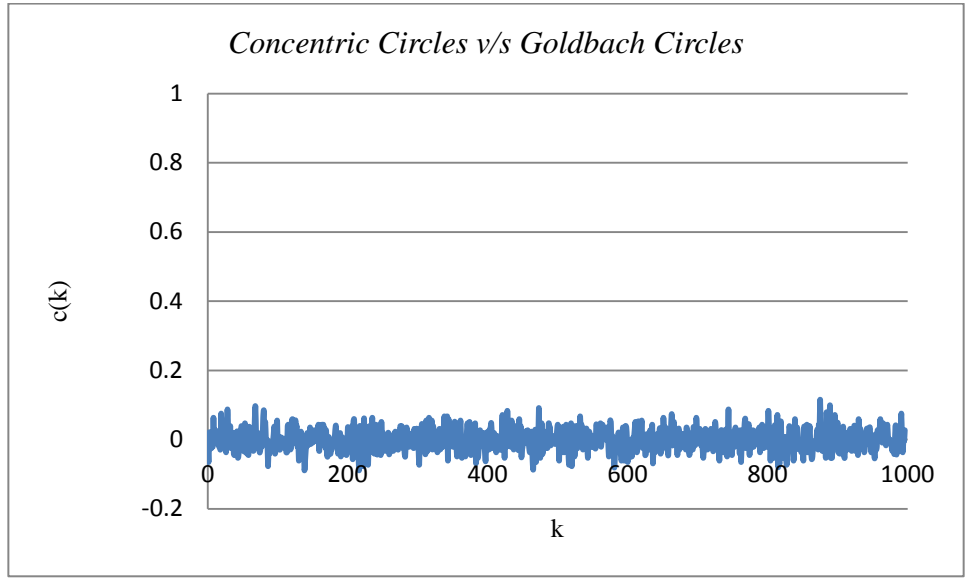

**Figure 7**. Cross correlation between concentric and Goldbach circle sequences

We find that the cross correlation values appear about similar in all the figures. This means that the cross correlation between ellipse and ellipse or ellipse and circle sequences has similar characteristics.

## CONCLUSIONS

This paper has presented numerical results on counts of concentric circles associated with an even number. This count has the same peaks as the count of Goldbach partitions. The number of circle sequences can be increased by sampling from each of the concentric circles of the numbers as long as certain conditions on the number of circles are satisfied.

We further show that binary circle and ellipse random sequences have excellent cross correlation properties with the peak value of less than five percent for relatively long sequences. For a good receiver the interference magnitude of five percent can be suppressed by the nonlinearity of the receiver processing. Since their autocorrelation properties are also excellent, they can be used in spread spectrum and other cryptographic applications.

**Acknowledgement.** This research was supported in part by research grant #1117068 from the National Science Foundation.